# Communication-robust and Privacy-safe Distributed Estimation for Heterogeneous Community-level Behind-the-meter Solar Power Generation

Jinglei Feng, *Student Member, IEEE*, Zhengshuo Li, *Senior Member, IEEE*

*Abstract*—**The rapid growth of behind-the-meter (BTM) solar power generation systems presents challenges for distribution system planning and scheduling due to invisible solar power generation. To address the data leakage problem of centralized machine-learning methods in BTM solar power generation estimation, federated learning (FL) method has been investigated for its distributed learning capability. However, conventional FL method has encountered various challenges, including heterogeneity, communication failures, and malicious privacy attacks. To overcome these challenges, this study proposes a communication-robust and privacy-safe distributed estimation method for heterogeneous community-level BTM solar power generation. Specifically, this study adopts multi-task FL as the main structure and learns the common and unique features of all communities. Simultaneously, it embeds an updated parameters estimation method into the multi-task FL, automatically identifies similarities between any two clients, and estimates the updated parameters for unavailable clients to mitigate the negative effects of communication failures. Finally, this study adopts a differential privacy mechanism under the dynamic privacy budget allocation strategy to combat malicious privacy attacks and improve model training efficiency. Case studies show that in the presence of heterogeneity and communication failures, the proposed method exhibits better estimation accuracy and convergence performance as compared with traditional FL and localized learning methods, while providing stronger privacy protection.**

## I. INTRODUCTION

DISTRIBUTED solar power generation has rapidly developed around the world. However, most small-capacity distributed solar power generation systems are installed behind-the-meter (BTM), and utilities have access only to the net load, which represents the actual load minus the solar power generation [1–4]. For example, in the United States, hardly any small-capacity solar power generation systems (< 10 kW) are equipped with direct measurement devices [5]. The invisibility of solar power generation may pose challenges for utilities, such as unacceptable forecasting errors for the net load, inaccurate reliability analysis, and hindrances to the design of optimal service restoration plans [6]. Therefore, exploring methods for accurately estimating BTM solar power generation is essential.

BTM solar power generation estimation methods can be classified into model-based and data-driven methods. Model-based methods infer the physical parameters of solar power generation systems based on smart meter data and then estimate solar power generation at any given time using the physical model of solar power generation systems [7–8]. However, simulating the effects of weather types, cloud motion, shading, and pollution on solar modules is difficult for model-based methods and, accordingly, these methods are unable to achieve satisfactory accuracy [9].

Data-driven methods are often employed to estimate BTM solar power generation based on smart meter data and weather information without relying on the physical model of solar power generation systems. Common data-driven methods include context-supervised source separation methods [10–12], dictionary learning methods [13–15], and machine-learning methods [16–17]. Of these, machine-learning methods such as decision trees, random forest and neural network have been applied in BTM solar power generation estimation and have achieved good performance. Because of the significant effects of data quality and quantity on the performance of machine-learning methods, these methods typically adopt a centralized approach. This requires that smart meter data from all communities be uploaded to a central service platform for training purposes.

With the promulgation of the European Union (EU) General Data Protection Regulations, more attention is being given to privacy security. The centralized method may lead to privacy issues, such as the risk of leakage of information related to community electricity usage patterns and locations. Researchers have focused on distributed estimation methods that retain all data locally. Of these methods, federated learning (FL) method has been widely utilized in areas such as load forecasting [18] and non-intrusive load monitoring [19] due to its distributed learning capability and has been proven to be a viable solution. Through FL, each client trains the machine-learning model on local device and uploads the model parameters to the server without sharing raw data. This reduces the risk of information leakage.

Although FL has shown promise in distributed estimation for community-level BTM solar power generation, several challenges remain must be addressed:

This work was supported by National Key R&D Program of China under Grant 2022YFB2402900. (Corresponding author: Zhengshuo Li).

The authors are with the School of Electrical Engineering, Shandong University, Jinan 250061, China. (e-mail: zsli@sdu.edu.cn).

1) *Heterogeneity*. Due to differences in meteorological conditions and solar power generation system configurations, each community has unique solar power generation characteristics. Conventional FL trains only a single global model and may not adapt well to the specific needs of each community.

2) *Communication failures*. Communication failures often occur during the FL process due to unstable communication links or network blockages. This results in the appearance of *unavailable clients* who are unable to upload their updated parameters to the server in a timely manner. The lack of updated parameter information from *unavailable clients* may lead to a decrease in global model performance.

3) *Privacy safety*. FL method uploads only the model parameters and not the raw data, thereby providing stronger privacy protection than centralized method. However, [20–21] indicate that attackers may still be able to restore sensitive information through model parameters and infer whether the obtained data comes from a specific client. This can result in user electricity data leakage and huge penalties for power companies.

Personalized FL methods are effective in addressing the problem of heterogeneity. Ref. [22] adopts a layer-wise parameters aggregation strategy to obtain personalized models for each community. However, the hyperparameters optimization process is relatively cumbersome. Ref. [23] proposes a multi-task FL framework, *Ditto*, which simultaneously trains both global and personalized models during the FL process to solve heterogeneous problems. However, issues remain in terms of designing an effective FL method that can address both communication failures and malicious privacy attacks.

To address these issues, this study proposes a communication-robust and privacy-safe distributed estimation method for heterogeneous community-level BTM solar power generation.

The primary contributions of this study are three-fold:

1) To address the significant community heterogeneity, this study introduces a multi-task FL framework [23] that embeds a *personalized model training task* in the local training process. The client can utilize a real-time updated global model and local data to train a personalized model that can fulfill the specific characteristics of each client. Case studies show that, the estimation error of the proposed method was reduced by 10% or more compared with the Fedavg method.

2) To compensate for the decrease in model performance caused by communication failures, this study adopts an updated parameters estimation method for unavailable clients [24]. When the communication process is normal, the server calculates the similarity between any two clients based on the updated parameters. When communication failures occur, the server uses the updated parameters of the client, which are most similar to those of the unavailable client to replace those of the unavailable client.

3) This study adopts a differential privacy mechanism under the dynamic privacy budget allocation strategy [25] to effectively address malicious privacy attacks and improve the model training efficiency when communication failures occur in FL process. Specifically, when communication failure occurs in a certain round, the privacy budget for subsequent rounds is dynamically adjusted to avoid wasting the privacy budget.

The remainder of this paper is structured as follows. Section II introduces the problems of applying conventional FL method to community-level BTM solar power generation estimation. Section III presents the implementation details of the proposed method. Section IV describes case studies that were conducted using publicly available datasets to validate the effectiveness of the proposed method. Finally, Section V offers the conclusions.

## II. PROBLEM STATEMENT

### A. Physical Scene

Fig. 1 shows the physical scene for the community-level BTM solar power generation estimation. A 10-kV substation is connected to multiple 0.4-kV secondary substations, each of which can be regarded as an independent community. The power distribution room converts voltage from 10 kV to 0.4 kV to supply users within the community. To simplify the analysis, we only consider solar power generation users in each community, as utilities typically have relevant information about the deployment of solar power generation systems, making it easy to distinguish solar power generation users from ordinary users.

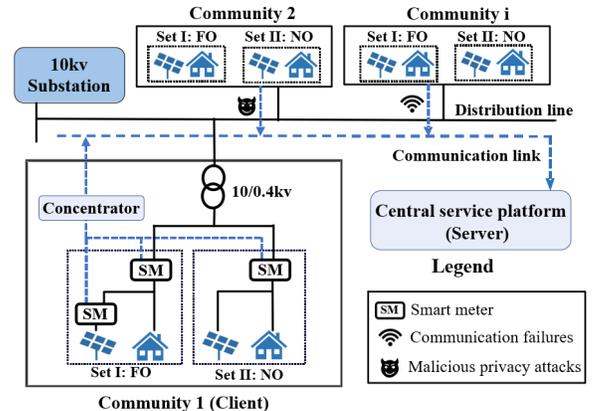

**Fig. 1.** Physical scene for the community-level BTM solar power generation estimation.

In each community, solar power generation users can be divided into two sets based on whether separate measurements are conducted for solar power generation.

*Set I*: Set of observable solar power generation users, denoted as FO, whose solar power generation is measured separately using smart meters. Therefore, the solar power generation, load, and net load of FO are all known.

*Set II*: Set of unobservable solar power generation users, denoted as NO, who only measures the net load. Therefore, the solar power generation of NO is unknown.

The local data concentrator collects data from the smart meters, processes it and uploads it to the central service platform via the public network. Communication failures and malicious privacy attacks may occur during data transmission.

### B. Community-level BTM Solar Power Generation Estimation based on a Deep Neural Network

The goal of community-level BTM solar power generation estimation is to obtain the total solar power generation of the community, which is the sum of the solar power generation of FO and NO. One simple and easy-to-apply method involves training a deep neural network (DNN) using historical data. For the FO in the $i$-th community, the meteorological data, net load and solar power generation are all known. Therefore, BTM solar power generation estimation can be regarded as a supervised machine-learning task. Using the meteorological data and net load as input features and solar power generation as the target, a DNN model is trained, as shown below,

$$PV_{FO,i} = f_{FO,i}(w_{FO,i}, T_{FO,i}, Net_{FO,i}), \qquad (1)$$

where $f_{FO,i}$ is the estimation model of solar power generation for FO in the $i$-th community; $T_{FO,i}$ denotes the set of meteorological data for FO in the $i$-th community; $PV_{FO,i}$ and $Net_{FO,i}$ represent the solar power generation and net load of FO in the $i$-th community, respectively; and $w_{FO,i}$ denotes the model parameters of the DNN for FO in the $i$-th community and includes the weights and biases of the neurons in each layer.

Because the two sets of solar power generation users (FO and NO) in the same community have similar meteorological conditions and solar power generation system configurations, they are likely to have similar solar power generation characteristics. Therefore, this study utilizes the DNN model $f_{FO,i}$ to approximate the total solar power generation model $f_i$ for the $i$-th community [22].

### C. FL Method

In order to utilize information from multiple communities while preventing data leakage, the FL method is adopted to train the DNN estimation model. The central service platform can be regarded as a *server* and each community can be regarded as a *client*.

Fig. 2 shows a comparison of the processes of the centralized and FL methods. In the centralized method, all communities upload local data and the central service platform trains the DNN model. However, in the FL method, each client trains the DNN model based on local data and the server uses the model parameters from the clients to update the global model.

In the FL method, the loss function of the DNN model for client $i$ is shown in (2), where $x_{i,k}$ and $y_{i,k}$ are the input features and the target of the $k$-th training sample for client $i$, respectively. In addition, $w$ is the DNN model parameter and $|D_i|$ is the number of training samples for client $i$. The global goal of FL is to minimize the average estimation error among the participating clients, as shown in (3), where $N$ denotes the number of clients, $|D|$ denotes the number of training samples for all clients, and $F_i(w)$ denotes the loss function for client $i$.

$$min_w F_i(w) = \frac{1}{|D_i|} \sum_{k=1}^{|D_i|} (f(w; x_{i,k}) - y_{i,k})^2 \qquad (2)$$

$$min_w G(w) = \sum_{i=1}^{N} \frac{|D_i|}{|D|} F_i(w) \qquad (3)$$

The entire FL training process can be divided into multiple communication rounds. Each round of FL consists of four main steps, as shown in Fig. 2(b):

1) the server broadcasts the updated global model;
2) the client trains the local model based on local data;
3) the client uploads local model parameters to the server;
4) the server updates the global model.

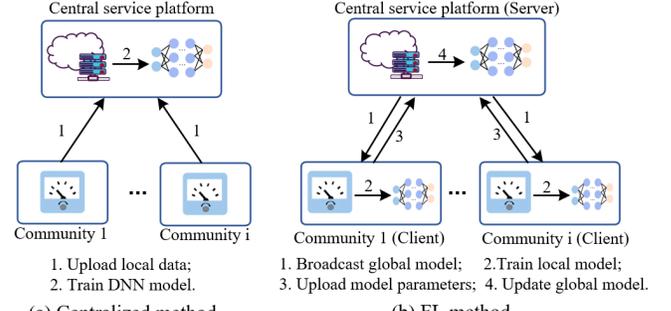

**Fig. 2.** Comparison of processes between the centralized and FL methods.

However, when the FL method is applied to the community-level BTM solar power generation estimation as described in Section II-A and II-B, some technical issues remain. First, in the presence of the community heterogeneity, the global model trained using the FL method cannot meet the needs of each community. Second, due to communication failures, some clients are unable to timely upload updated parameters to the server, which results in a decrease in the global model performance. In addition, malicious privacy attacks may occur during the communication process of the FL method.

## III. METHODOLOGY

### A. Overview of the Proposed Method

To address the issues of community heterogeneity, communication failures, and malicious privacy attacks, this study proposes a communication-robust and privacy-safe distributed estimation method for heterogeneous community-level BTM solar power generation. Fig. 3 illustrates the structure of the proposed method.

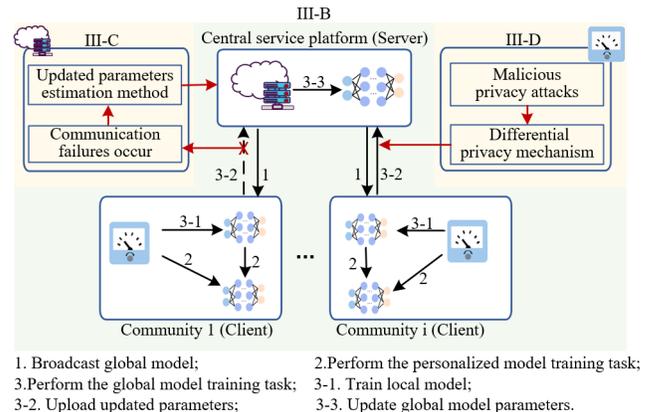

**Fig. 3.** Structure of the proposed method.

Heterogeneity is a major challenge in community-level BTM solar power generation distributed estimation. This study adopts multi-task FL [23] as the main framework to train the DNN estimation model, as shown by the cyan shading in Fig. 3. In each communication round, multi-task FL simultaneously performs the following two tasks: a *personalized model training task* is executed by each client using global model parameters and local data to update the personalized model, and a *global model training task* is jointly executed by the client and server by aggregating the local model parameters from the clients to update the global model. Multi-task FL is described in detail in Section III-B.

Moreover, we notice that communication failures may occur during the communication process of the *global model training task* (i.e., Step 3-2 in Fig. 3), which will negatively affect the performance of the DNN model. To resolve this issue, we further embed an updated parameters estimation method [24] in Step 3-2. When communication failures occur, the server automatically estimates the updated parameters of *unavailable clients* based on those of other *available clients*, the details of which are described in Section III-C.

In addition, to prevent malicious attackers from restoring sensitive information through model parameters, this study also embeds a differential privacy mechanism under the dynamic privacy budget allocation strategy [25] in Step 3-2. The client adds noise to the model parameters and automatically adjusts the privacy budget for subsequent rounds when communication failures occur. The details are described in Section III-D.

### B. Multi-task FL Framework

Compared with the conventional FL method as described in Section II-C, the multi-task FL method essentially embeds a *personalized model training task* in the local training process, as shown in (4),

$$min_{v_i} L_i(v_i, w^*) = F_i(v_i) + \frac{\mu}{2}\|v_i - w^*\|^2, \qquad (4)$$
$$s.t. \, w^* \in argmin_w G(w)$$

where $v_i$ denotes the personalized model parameters for client $i$, $w^*$ denotes the global model parameters, $\mu$ is the regularization factor that controls the trade-off between the personalized and global models. When regularization factor $\mu$ is set to 0, it is simplified to train the local model. As the regularization factor $\mu$ increases, the personalized model gradually approaches the global model.

By utilizing local data and global model information for joint training, personalized models can learn the unique features of each community and the common features of all communities, thereby improving estimation accuracy.

The process of the multi-task FL method is indicated by the cyan shading in Fig. 3, and the steps are described below.

**Step 1:** Broadcast global model. The server broadcasts the latest global model parameters to each client. During the first round of training, the server randomly initializes the global model parameters and determines the training parameters of the entire FL process.

**Step 2:** The personalized model training task is performed. The personalized model parameters in the latest round are used as the initial values for the current round (during the first round of training, the client randomly initializes the personalized model parameters). The client then performs multiple iterations of stochastic gradient descent (SGD) to update the personalized model parameters. The optimization objective and SGD process are given by the following (5) and (6), respectively,

$$min_{v_i^r} L_i(v_i^r, w^r) = F_i(v_i^r) + \frac{\mu}{2}\|v_i^r - w^r\|^2, \qquad (5)$$
$$v_i^r \leftarrow v_i^r - \eta_1 \nabla L_i(v_i^r, \ w^r), \qquad (6)$$

where $w^r$ denotes the global model parameters in the $r$-th round, $v_i^r$ is the personalized model parameter for client $i$ in the $r$-th round, $\eta_1$ is the learning rate during the SGD process, and $\nabla L_i$ is the gradient of the function $L_i$. The number of the iteration process of SGD in Step 2 is defined as $E_1$.

**Step 3:** The global model training task is performed. This includes three steps: training the local model, uploading the updated parameters, and updating the global model parameters.

**Step 3-1:** The local model is trained. Each client uses the latest global model parameters as the initial values and then performs multiple iteration process of SGD to update the local model parameters based on local data. The optimization objective and SGD process are given by the following (7) and (8), respectively.

$$min_{w_i^r} F_i(w_i^r) = \frac{1}{|D_i|}\sum_{k=1}^{|D_i|}(f(w_i^r; x_{i,k}) - y_{i,k})^2, \quad (7)$$
$$w_i^r \leftarrow w_i^r - \eta_2 \nabla F_i(w_i^r), \qquad (8)$$

where $w_i^r$ denotes the local model parameters for client $i$ in the $r$-th round and is used to update the global model parameters. In addition, $\eta_2$ is the learning rate during the SGD process. The number of the iteration process of SGD in this step is defined as $E_2$.

**Step 3-2:** The updated parameters are uploaded. Each client uploads the updated parameters of the local model to the server after completing local model training, as shown by equation (9),

$$\Delta w_i^r = w_i^r - w^r, \qquad (9)$$

where $w^r$ denotes the global model parameters in the $r$-th round, $w_i^r$ denotes the local model parameters for client $i$ in the $r$-th round, and $\Delta w_i^r$ denotes the updated parameters of the local model for client $i$ in the $r$-th round.

**Step 3-3:** The global model parameters are updated. After the updated parameters of all clients are uploaded, the server updates the global model parameters, as shown in (10),

$$w^{r+1} = w^r + \sum_{i=1}^{N} \frac{|D_i|}{|D|}\Delta w_i^r, \qquad (10)$$

where $w^{r+1}$ denotes the global model parameters in the (r+1)-th round.

Steps 1–3 are repeated until the model parameters converge or the specific number of communication rounds is reached. The final $v_i$ represents the desired personalized model parameters.

## C. Updated Parameters Estimation for Unavailable Clients

In the *global model training task*, we assume that all clients are involved in each communication round. However, due to communication failures, some clients are unable to upload their updated parameters to the server in a timely manner, which will cause a decrease in global model performance. When communication failures are severe, the global model may fail to converge [26].

This study adopts an updated parameters estimation method [24] to address communication failures. Before the global model parameters are updated (i.e., Step 3-3) the server *calculates the similarity score* between any two available clients and *estimates the updated parameters for unavailable clients*.

***Calculate similarity score:*** Let us assume that in the *r*-th round, $U^r$ and $A^r$ denotes the set of unavailable clients and the set of available clients, respectively. The server can then obtain the updated parameters of all clients in $A^r$. During the local model training process, the initial model parameters are the same. Therefore, when the updated parameters of any two clients are similar, their solar power generation characteristics are also similar. It is also more likely that these two clients will have similar updated parameters in the subsequent rounds. This enables using the updated parameters of similarly available clients to replace those of unavailable clients.

In the proposed method, the normalized cosine distance of the updated parameters is used as a measure of the similarity score. The calculation of the similarity score between any two clients in the *r*-th round is shown in (11),

$$s_{i,j}^r = \frac{1}{2}(\frac{\Delta w_i^r \cdot \Delta w_j^r}{\|\Delta w_i^r\|\|\Delta w_j^r\|} + 1), \tag{11}$$

where $\Delta w_i^r \cdot \Delta w_j^r$ is the vector product of $\Delta w_i^r$ and $\Delta w_j^r$, and $\|\cdot\|$ is the amplitude of the vector. The range of similarity score $s_{i,j}^r$ is [0,1]. When the similarity score between two clients approaches 1, the updated parameters of these two clients are more similar.

The normalized cosine distance is chosen as the similarity score metric instead of the Euclidean distance because, during the model parameters update process, the focus is on the direction of the updated parameters, and the size of the updated parameters is not critical.

However, a similarity score calculated in a single round does not provide accurate similarity information. Thus, the server adopts an average similarity score $S_{i,j}^r$ to measure the similarity between clients $i$ and $j$, as shown in (12),

$$S_{i,j}^r = \begin{cases} \frac{N_{i,j}^{r-1}}{N_{i,j}^{r-1}+1}S_{i,j}^{r-1} + \frac{1}{N_{i,j}^{r-1}+1}s_{i,j}^r, & if \ i,j \in A^r \\ S_{i,j}^{r-1}, & otherwise \end{cases} \tag{12}$$

where $N_{i,j}^{r-1}$ is the number of rounds in which clients $i$ and $j$ are available before the *r*-th round.

***Estimate the updated parameters for unavailable clients:*** When unavailable client $i$ appears, that is, when the server has not received the updated parameters of client $i$ within the specified time, the server uses the updated parameters of client

$j$, who is most similar to *client i*, to replace the updated parameters of *unavailable client i*, as shown in (13),

$$\Delta \hat{w}_i^r \leftarrow \Delta w_j^r, \tag{13}$$

where $\Delta w_j^r$ denotes the updated parameters in the *r*-th round for client $j$, and $\Delta \hat{w}_i^r$ denotes the estimated parameters in the *r*-th round for unavailable client $i$.

## D. Differential Privacy Mechanism under the Dynamic Privacy Budget Allocation Strategy

During the communication process of *global model training task*, malicious privacy attacks may occur. This study adopts $\varepsilon$-differential privacy mechanism to enhance privacy protection by adding noise to the updated parameters. The $\varepsilon$-differential privacy is defined as follows.

**Definition 1** ($\varepsilon$-differential privacy) [27]. For a stochastic algorithm M and a dataset D, $P_M$ is the set of all output results when algorithm M is applied to dataset D. If for any subset $S \in P_M$ and any pair of adjacent datasets D and $D'$, algorithm M satisfies

$$\Pr[M(D) \in S] \le e^\varepsilon \cdot \Pr[M(D') \in S], \tag{14}$$

it can be said that algorithm M satisfies the $\varepsilon$-differential privacy. Here, $\varepsilon$ is the privacy budget, an index for measuring the degree of privacy protection. When $\varepsilon \to 0$, the probability of two adjacent datasets outputting the same result is similar. As a result, the attacker cannot infer whether the training samples come from a specific client. Thus, the purpose of privacy protection is achieved.

The $\varepsilon$-differential privacy mechanism in FL mainly involves local model parameters doing the actions of clipping and adding noise. The updated parameters for clipping and adding noise of client $i$ in the *r*-th round are as follows.

$$\Delta \bar{w}_i^r = \frac{\Delta w_i^r}{\max\left(1, \frac{\|\Delta w_i^r\|}{C}\right)}, \tag{15}$$

$$\Delta \tilde{w}_i^r = \Delta \bar{w}_i^r + n_D, \tag{16}$$

where $\Delta \bar{w}_i^r$ and $\Delta \tilde{w}_i^r$ represent the updated parameters after clipping and the updated parameters after adding noise, respectively; C is the clipping threshold, with the aim of limiting updated parameters $\Delta w_i^r$ to $(-C, C)$; and $n_D$ is the stochastic noise generated by each client via the Laplace mechanism, as given by the following equation (17).

$$Laplace(\mu, b) = \frac{1}{2b}\exp\left(-\left(\frac{|x-\mu|}{b}\right)\right) \tag{17}$$

The added stochastic noise satisfies the Laplace distribution and $\mu = 0$, b=$\Delta s / \varepsilon$. $\varepsilon$ is the privacy budget and $\Delta s$ is the sensitivity, as given by the following equation (18). Finally, the updated parameters are uploaded after adding noise.

$$\Delta s = \frac{2C}{|D_i|}. \tag{18}$$

The most important parameter is the privacy budget $\varepsilon$. A smaller $\varepsilon$ corresponds to stronger noise and to stronger privacy protection, but the efficiency of training decreases. Investigation into the effect of privacy budget $\varepsilon$ is performed in Section IV-D.

In the conventional differential privacy mechanism, each client is assigned a fixed privacy budget for each round of communication. The differential privacy mechanism has sequential composition property as shown in Theorem 1. For example, if the client is assigned privacy budget $\varepsilon$ for each round and performs R rounds of communication, then the total privacy budget is $R\varepsilon$.

**Theorem 1** (Sequential composition property) [27]. Given dataset $D$ and a set of differential privacy algorithms $M_1$, $M_2$, and $M_i$ related to $D$, algorithm $M_i(D)$ satisfies the $\varepsilon_i$ - differential privacy and the random processes of any two algorithms are independent. The sequence of these algorithms satisfies $\bar{\varepsilon}$-differential privacy, where $\bar{\varepsilon} = \sum_{i=1}^{m} \varepsilon_i$.

**Proof** [27]. For any sequence $r$ of outcomes $r_i \in Range(M_i(D))$, we write $M_i^r$ for algorithm $M_i$ supplied with $r_1, \cdots, r_{i-1}$. The probability of output r from the sequence of $M_i(D)$ is

$$\Pr[M(D) = r] = \prod_i \Pr[M_i^r(D) = r_i] \quad (19)$$

Correspondingly, we can state that

$$\Pr[M(D') = r] = \prod_i \Pr[M_i^r(D') = r_i] \quad (20)$$

Applying the definition of differential privacy for each $M_i^r$, we obtain

$$\prod_i \Pr[M_i^r(D) = r_i]$$
$$\leq \prod_i \Pr[M_i^r(D') = r_i] \times \prod_i e^{\varepsilon_i} \quad (21)$$

By substituting (19) and (20) into (21), we prove Theorem 1.

However, due to communication failures, the client is unable to upload the updated parameters in certain rounds, which means that the privacy budget in these rounds is wasted.

This study adopts a dynamic privacy budget allocation strategy [25] that allocates wasted privacy budgets to future training rounds to improve the training efficiency of the model. Specifically, when communication failures occur, because the overall privacy budget remains unchanged, the privacy budget for all subsequent rounds must be reallocated.

At the beginning of training, the privacy budget for each round is $\varepsilon^0$ and R communication rounds are required. When a communication failure occurs in round $r_1$, the privacy budget for this round is saved, and the privacy budget for all subsequent rounds becomes $\varepsilon^1$, as given by the following equation.

$$\varepsilon^1 = \frac{R - r_1 + 1}{R - r_1} \varepsilon^0. \quad (22)$$

When the communication problem occurs in the round $r_i$, the original privacy budget for the current round is $\varepsilon^{i-1}$ and the privacy budget for subsequent rounds becomes $\varepsilon^i$, as shown in (23). Fig. 4 shows the adjustment process for the privacy budget.

$$\varepsilon^i = \frac{R - r_i + 1}{R - r_i} \varepsilon^{i-1}. \quad (23)$$

The dynamic privacy budget allocation strategy adopted in this study can allocate the wasted privacy budget due to communication failures to subsequent rounds without changing the total privacy budget, thereby improving the training efficiency of the model.

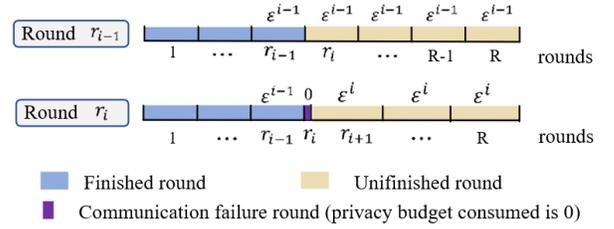

**Fig. 4.** Simple illustration of the privacy budget allocation strategy.

### E. Overall Algorithm

To illustrate the specific work of the server and client in the proposed method, Fig. 5 shows the flow of the proposed method. The pseudocode for the proposed method is presented in Algorithm 1.

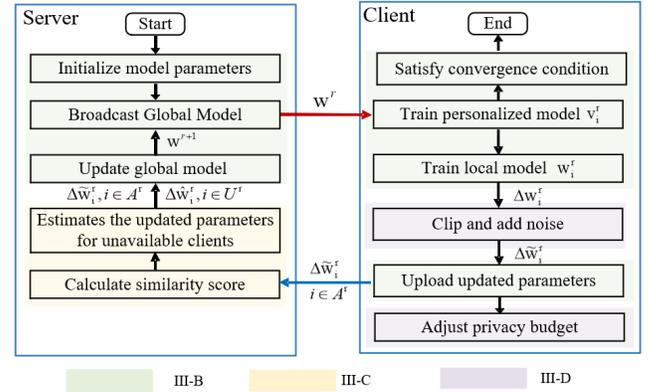

**Fig. 5.** Flow of the proposed method.

---

**Algorithm 1** Pseudocode of the Proposed Communication-robust and Privacy-safe Distributed Estimation Method

**Input:** The set of clients is $C$; number of clients is $N$; number of samples for client $i$ is $|D_i|$; number of communication rounds is $R$; numbers of training epochs are $E_1, E_2$; the learning rates are $\eta_1, \eta_2$; the regularization factor is $\mu$.

**Output:** Personalized model $v_i$ for each client.

1  Server initializes global model parameters
2  **for** $r = 0, \cdots, R - 1$ **do**
3      Server broadcasts global model parameters $w^r$ to each client
4      **for** $i = 1, \cdots, N$ **do**
5          Update personalized model $v_i^r$ for $E_1$ local iterations on $L_i$:
        $v_i^r \leftarrow v_i^r - \eta_1 \nabla L_i(v_i^r, \ w^r)$
6          Update local model $w_i^r$ for $E_2$ local iterations on $F_i$:
        $w_i^r \leftarrow w_i^r - \eta_2 \nabla F_i(w_i^r)$
7          Calculate updated parameters: $\Delta w_i^r = w_i^r - w^r$
8          Clip updated parameters:
        $\Delta \tilde{w}_i^r = \Delta w_i^r / \max\left(1, \frac{\|\Delta w_i^r\|}{C}\right)$
9          Add noise: $\Delta \bar{w}_i^r = \Delta \tilde{w}_i^r + n_D$
10         Client uploads $\Delta \bar{w}_i^r$ to server
11         Client adjusts privacy budget based on communication condition
    **end**
12     Server calculates the similarity score between any two available clients and estimates the updated parameters for unavailable clients
13     Server updates the next round of global model parameters: $w^{r+1} = w^r + \sum_{i=1}^{N} \frac{|D_i|}{|D|} \Delta w_i^r$
**end**

## A. Experimental Settings

*1) Dataset:* In the experiment, smart meter data were obtained from the available energy dataset of the Australian grid company Ausgrid [28]. This dataset records the solar power generation, load, and net load of 300 solar power generation users every 30 min. Meteorological data were provided by the National Solar Radiation Database (NSRDB) [29] and included irradiance, temperature, relative humidity and wind speed, with a spatial resolution of $4 \times 4$ kilometers and a time resolution of 30 min.

To model the problem of BTM solar power generation estimation in heterogeneous communities, k-means clustering was performed on solar power generation users based on their latitude and longitude positions, thus dividing 300 solar power generation users into k communities. In each community, 60% and 40% of users were selected as FO and NO users, respectively. The electricity data of FO users in the community were gathered as the training set, and the electricity data of all users in the community were gathered as the testing set.

*2) Parameter Settings:* This study used a DNN as the estimation model. The input features included net load, irradiance, temperature, relative humidity, and wind speed [22]. The output feature was the solar power generation of the community. Table I lists the structural parameters of the DNN and the training parameters of the FL.

TABLE I
PARAMETERS USED IN THE EXPERIMENT

| Parameter | Value | | |
|---|---|---|---|
| Number of hidden layers | 1 | | |
| Number of hidden layer neurons | 40 | | |
| Activation function | ReLU | | |
| Parameter | Value | Parameter | Value |
| $R$ | 200 | $\mu$ | 5e-4 |
| $E_1$ | 5 | $E_2$ | 10 |
| $\eta_1$ | 0.01 | $\eta_2$ | 0.01 |

To simulate the scenario of communication failures, during each round of communication, we randomly selected some clients as *unavailable clients*. For ease of description, we use $n_f$ to denote the maximum number of unavailable clients per round. In addition, $n_c$ represents the ratio of the maximum number of unavailable clients to the total number of clients. When $n_c = 0$, no communication failure occurs during the FL process. A larger $n_c$ corresponds to a more severe communication failure.

*3) Comparison Methods and Performance Evaluation:* The following four comparison methods were conducted to verify the effectiveness of the proposed method.

*A1*: Localized learning method: For each community, only their own data were used to train the localized models.

*A2*: Fedavg method [30]: Each community trains models on local devices and then uploads the model parameters to a central service platform.

*A3*: This method includes multi-task FL, updated parameters estimation for unavailable clients, and a conventional differential privacy mechanism [31].

*A4*: The proposed method includes multi-task FL, updated parameters estimation for unavailable clients, and a differential privacy mechanism under the dynamic privacy budget allocation strategy.

The difference between methods A3 and A4 is that the differential privacy mechanism of the A4 method adopts the dynamic privacy budget allocation strategy. Therefore, under conditions of community heterogeneity, this study compared the performance of methods A1, A2, and A4 to verify the effectiveness of the proposed method in addressing heterogeneity. Because the A1 method operates only locally, in a communication failure scenario, this study compared the performances of methods A2 and A4 to demonstrate the advantages of the proposed method in dealing with communication failures. Finally, we compared methods A3 and A4 to demonstrate that the dynamic privacy budget allocation strategy adopted in this study could improve the model training efficiency while maintaining the overall level of privacy protection.

This study leveraged the normalized root-mean-square error (NRMSE) [19] as the evaluation indicator of the model estimation performance, which is calculated by the following equation,

$$NRMSE = \frac{\sqrt{\frac{1}{m} \sum_{i=1}^{m} \left(\widehat{Y_i} - Y_i\right)^2}}{Y_{max} - Y_{min}}, \qquad (24)$$

where $m$ denotes the number of samples, $Y_i$ and $\widehat{Y_i}$ are the ground truth and the estimated values of sample $i$, respectively, and $Y_{max}$ and $Y_{min}$ are the maximum and minimum values among all samples, respectively. The NRMSE can be used to compare the error levels of different datasets. A smaller NRMSE corresponds to a higher estimation accuracy.

*4) Code Implement:* The code was implemented in Python, and the deep learning framework was PyTorch.

## B. Performance of the Multi-task FL

We next demonstrate the performance advantages of the proposed method as compared with other methods when dealing with community heterogeneity and provide a reference for the selection of parameters with the proposed method. Here, we consider four communities.

*1) Comparison with Other Methods:* Table II lists the NRMSE values of the three methods (A1, A2, and A4) across all communities. The heterogeneity of different communities is reflected in the weather conditions and the configuration of solar power generation systems. In the following, "community" is abbreviated to "com."

From Table II, the following observations can be made: (a) Among the three methods, the A1 method typically exhibited the worst performance because only limited data were used for training, and the estimated performance for non-occurring weather events was poor. (b) When using the A2 method, the estimation performance of Community 2 and 3 was poor, indicating that the global model trained by the Fedavg method cannot achieve good results on all clients. (c) The estimated performance of A4 method was the best across all communities, and the estimation error was reduced by 10% or more as compared with the A2 method. This was because the A4 method simultaneously performs personalized and global model training tasks. It can access all community model parameters and cope with the effects of various weather conditions on solar power generation. In addition, the personalized model training task captures the configuration of solar power generation systems in each community.

TABLE II
NRMSE VALUES AMONG THREE METHODS
ACROSS HETEROGENEOUS COMMUNITIES

| NRMSE | Com 1 | Com 2 | Com 3 | Com 4 |
|---|---|---|---|---|
| A1 | 0.144 | 0.158 | 0.148 | 0.169 |
| A2 | 0.118 | 0.144 | 0.143 | 0.115 |
| A4(proposed) | **0.099** | **0.115** | **0.115** | **0.103** |

2) *Effects of Parameters:* Fig. 6 shows the NRMSE values under different numbers of personalized model training epochs $E_1$, and different numbers of local model training epochs $E_2$. As $E_2$ increased, the estimated error of the A4 method showed a decreasing trend. After $E_2$ exceeded 6, the estimated error of the A4 method remains unchanged. This indicated that when choosing $E_2$, it should be slightly more than $E_1$. In this scenario, more personalized characteristics of the community can be learned, thereby improving the performance of the model.

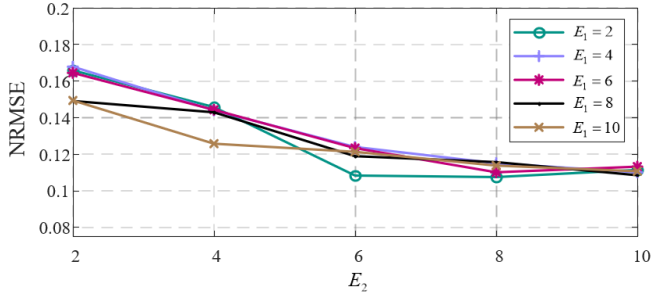

**Fig. 6.** NRMSE values with different $E_1$ and $E_2$ for community-3 using the A4 method.

Table III shows the NRMSE values for the A4 method under different regularization factors $\mu$. The table shows that the NRMSE reaches the minimum value when $\mu$ = 5e-4. Here, $\mu$ was used to balance the learning of global and personalized characteristics. When the parameter $\mu$ was too large or too small, it caused a decrease in the performance of the proposed model. In addition, when the regularization factor was between 1e-4 and 1e-3, the performance of the A4 method was superior to that of both the A1 and A2 methods.

TABLE III
NRMSE VALUES WITH DIFFERENT $\mu$ UNDER THE A4 METHOD.

| | NRMSE | Com 1 | Com 2 | Com 3 | Com 4 |
|---|---|---|---|---|---|
| A4 | $\mu$=1e-4 | 0.110 | 0.134 | 0.132 | 0.112 |
| | $\mu$=5e-4 | **0.099** | **0.115** | **0.115** | **0.103** |
| | $\mu$=1e-3 | 0.103 | 0.128 | 0.129 | 0.106 |
| | $\mu$=5e-3 | 0.118 | 0.142 | 0.141 | 0.116 |

In summary, for the problem of community heterogeneity, the proposed method demonstrates the advantage of higher estimation accuracy as compared with the Fedavg and localized learning methods.

### C. Performance of the Updated parameters estimation method

We next examine the effects of communication failures on the estimated performance of the A2 method as well as the advantage of the proposed method in coping with communication failures. The proposed method uses the updated parameters of similar clients to estimate those of unavailable clients, making it suitable for situations involving several clients. Here, we consider 16 communities.

TABLE IV
NRMSE VALUES BETWEEN THE A2 AND A4 METHODS ACROSS
HETEROGENEOUS COMMUNITIES UNDER COMMUNICATION FAILURES

| NRMSE | $n_c = 0.25$ | | $n_c = 0.50$ | | $n_c = 0.75$ | |
|---|---|---|---|---|---|---|
| | A2 | A4 | A2 | A4 | A2 | A4 |
| Com 1 | 0.108 | 0.098 | 0.106 | 0.098 | 0.115 | 0.098 |
| Com 2 | 0.090 | 0.082 | 0.099 | 0.083 | 0.102 | 0.083 |
| Com 3 | 0.090 | 0.082 | 0.094 | 0.082 | 0.096 | 0.082 |
| Com 4 | 0.086 | 0.084 | 0.088 | 0.084 | 0.092 | 0.084 |
| Com 5 | 0.138 | 0.140 | 0.153 | 0.140 | 0.151 | 0.140 |
| Com 6 | 0.114 | 0.106 | 0.117 | 0.108 | 0.132 | 0.106 |
| Com 7 | 0.112 | 0.110 | 0.123 | 0.110 | 0.119 | 0.110 |
| Com 8 | 0.103 | 0.102 | 0.108 | 0.102 | 0.102 | 0.102 |
| Com 9 | 0.119 | 0.117 | 0.130 | 0.117 | 0.129 | 0.118 |
| Com 10 | 0.107 | 0.100 | 0.120 | 0.100 | 0.122 | 0.100 |
| Com 11 | 0.101 | 0.097 | 0.110 | 0.097 | 0.108 | 0.097 |
| Com 12 | 0.089 | 0.086 | 0.089 | 0.086 | 0.096 | 0.086 |
| Com 13 | 0.108 | 0.103 | 0.108 | 0.103 | 0.116 | 0.103 |
| Com 14 | 0.111 | 0.103 | 0.122 | 0.102 | 0.123 | 0.102 |
| Com 15 | 0.097 | 0.086 | 0.100 | 0.086 | 0.105 | 0.086 |
| Com 16 | 0.104 | 0.100 | 0.098 | 0.100 | 0.108 | 0.100 |

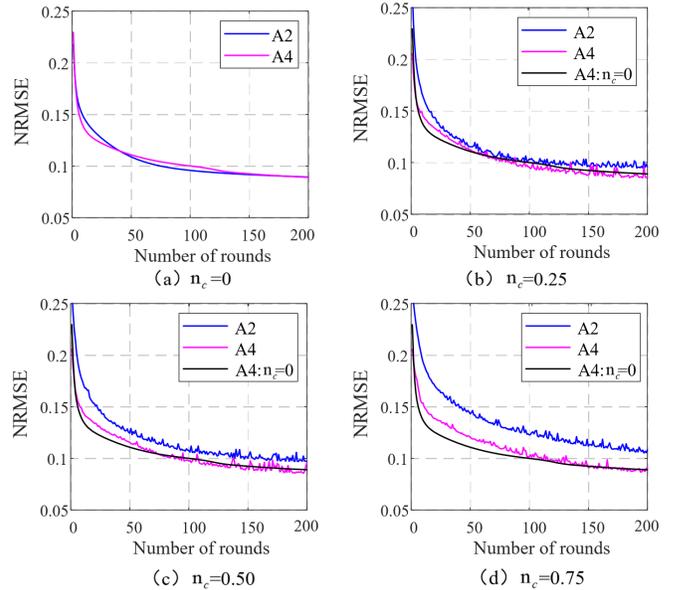

**Fig. 7.** Comparison of training curves between the A2 and A4 methods under communication failures.

Table IV shows a comparison of the NRMSE values between methods A2 and A4 with different $n_c$ under different communication failure scenarios. Fig. 7 shows the training curves of these two methods under different communication failure scenarios. The following conclusions were drawn.

First, as the communication failure scenario became more severe, the estimation error of the A2 method gradually increased in most communities. Compared to $n_c$ =0.25, the estimation error of the A2 method increased by 5%–15% at $n_c$ =

0.75. This was because the global model lost some useful information from unavailable clients.

Second, in the three communication failure scenarios, the estimation error of the A4 method remained basically unchanged in each community. On the one hand, the use of updated parameters estimation methods mitigates the global model performance degradation caused by communication failures. On the other hand, when a client encounters a communication failure in certain rounds, the personalized model training task continues, making the estimation model robust in dealing with communication failures.

Third, under different communication failure scenarios, the estimation accuracy and convergence performance of the A4 method were superior to those of the A2 method. In the case of $n_c = 0.75$, the estimation error of the A4 method was reduced by 10%–20% as compared with the A2 method across most communities.

Finally, when more severe communication failures occurred, the performance improvement of the A4 method was greater as compared with the A2 method.

From the aforementioned experiments, we showed that the proposed method is robust in dealing with communication failures and that its estimation accuracy across various communities and convergence performance are superior to those of the Fedavg method.

### D. Performance of the Differential Privacy Mechanism under the Dynamic Privacy Budget Allocation Strategy

We next consider the effect of privacy budget parameters on model performance and demonstrate the effectiveness of the proposed dynamic privacy budget allocation strategy.

*1) Effect of Privacy Budget $\varepsilon$:* Fig. 8 presents the NRMSE values of the A4 method under different privacy budget parameters without communication failures. Note that $\varepsilon$ is the privacy budget for each communication round. When $\varepsilon = \infty$, it indicates that there is no added noise. When $\varepsilon$ decreased from 1 to 0.005, the estimation performance of the model remained basically unchanged and was similar to the model performance without the added noise. This indicated that the noise generated by the differential privacy mechanism can be filtered out through the training process of DNN. Fig. 8 shows that when the privacy budget was less than 0.005, the estimation error of the model increased significantly. This indicated that the added noise was too large and negatively affected the estimation accuracy of the model. Therefore, we need to choose a $\varepsilon$ larger than 0.005 as an initial privacy budget in the experiment.

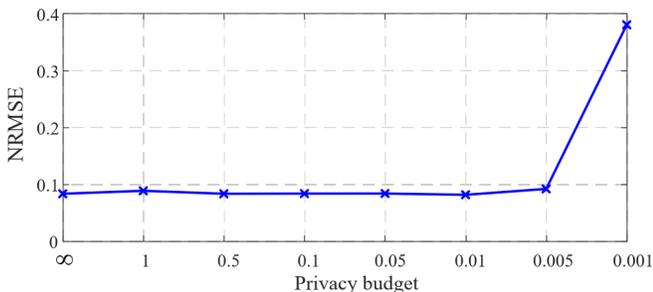

**Fig. 8.** NRMSE values under different privacy budget parameters for the A4 method.

*2) Effectiveness of the Dynamic Privacy Budget Allocation Strategy:* Fig. 9 shows a comparison of the training curves between the A3 and A4 models under different scenarios. Table V lists the NRMSE values between A3 and A4 models with different privacy budget in the case of communication failures.

In various scenarios, the convergence performance and estimation accuracy of the A4 method were improved as compared with those of the A3 method. The main reason for this was that under the dynamic privacy budget allocation strategy, the privacy budget that was saved from the communication failure rounds was allocated to subsequent rounds. With the overall level of privacy protection unchanged, an increase in the privacy budget for subsequent rounds indicated a decrease in added noise, thus improving the training efficiency of the model.

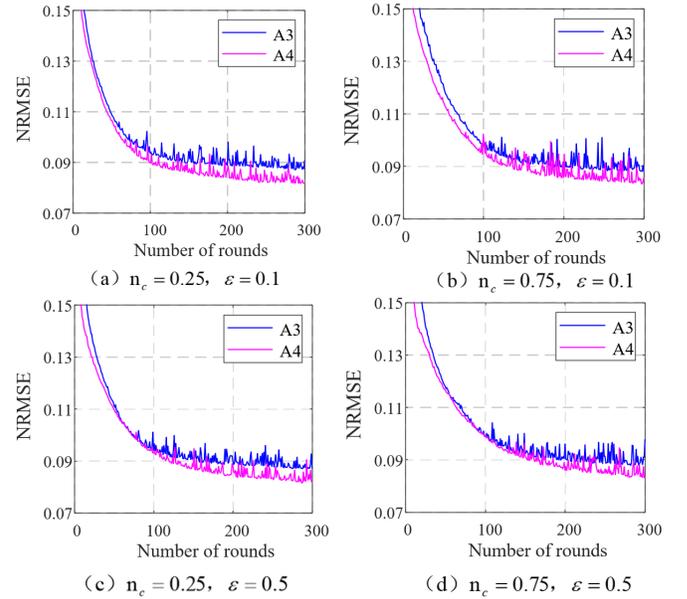

**Fig. 9.** Comparison of training curves between the A3 and A4 methods with different privacy budgets under communication failures.

**TABLE V**
**NRMSE VALUE BETWEEN A3 AND A4 WITH DIFFERENT PRIVACY BUDGET IN THE CASE OF COMMUNICATION FAILURES**

| NRMSE | $n_c = 0.25$ | | $n_c = 0.50$ | | $n_c = 0.75$ | |
|---|---|---|---|---|---|---|
| | A3 | A4 | A3 | A4 | A3 | A4 |
| $\varepsilon = 0.1$ | 0.087 | **0.081** | 0.088 | **0.082** | 0.089 | **0.084** |
| $\varepsilon = 0.5$ | 0.089 | **0.082** | 0.089 | **0.084** | 0.089 | **0.085** |
| $\varepsilon = 1$ | 0.090 | **0.084** | 0.088 | **0.084** | 0.091 | **0.086** |

In addition, the proposed method adopts a multi-task FL, in which the personalized model parameters are saved locally, which also plays a role in privacy protection to some extent.

## V. CONCLUSION

This study proposed a communication-robust and privacy-safe distributed estimation method for heterogeneous community-level BTM solar power generation. Through multi-task FL, each community's personalized model can be trained using real-time information from the global model and local data to fulfill the personalized characteristics of each community. The updated parameters estimation method was adopted to compensate for global model performance degradation caused by communication failures. In addition, a differential privacy mechanism with a dynamic privacy budget allocation strategy was adopted to add noise to the updated parameters, thereby providing a stronger level of privacy protection. The case studies confirmed that, compared with the Fedavg and localized learning methods, the proposed method exhibits better estimation accuracy under heterogeneous communities and communication failures. In addition, compared with the conventional differential privacy method, the proposed method improves the model training efficiency when communication failures occur in FL process.

With the rapid deployment of distributed solar power generation systems, invisible solar power generation at the community level significantly affects the scheduling and stability of power systems [32]. This method can assist power systems in achieving BTM solar power generation estimation, thereby achieving accurate power forecasting and reliability analysis.

Future research might include the application of transfer learning methods [33] to BTM solar power generation estimation to achieve rapid generation of BTM solar power generation estimation models under new scenarios.